\documentclass[english,12pt]{article}
\usepackage[cp1251]{inputenc}
\usepackage{babel}
\usepackage{amsfonts, amsmath}

\newcommand{\be}{\begin{equation}} \newcommand{\ee}{\end{equation}}

\def\A{\ensuremath{\boldsymbol{A}}}

\newcommand{\gsim}{\mathrel{\hbox{\rlap{\lower.55ex\hbox{$\sim$}} \kern-.3em \raise.4ex \hbox{$>$}}}}

\begin{document}
\begin{center}
{\bf Spacetime Quantum  Fluctuations, Minimal  Length  and  Einstein Equations}\\
\vspace{5mm} A.E.Shalyt-Margolin \footnote{E-mail:
a.shalyt@mail.ru; alexm@hep.by}\\ \vspace{5mm} \textit{National
Centre of Particles and High Energy Physics, Bogdanovich Str. 153,
Minsk 220040, Belarus}
\end{center}
\rm\normalsize \vspace{0.5cm}

\begin{abstract}
In the process of work it has been found that space-time quantum
fluctuations are naturally described in terms of the deformation
parameter introduced on going from the well-known quantum
mechanics to that at Planck’s scales and put forward in the
previous works of the author.  As shown, with the use of quite
natural assumptions, these fluctuations must be allowed for  in
Einstein Equations  to lead to the dependence of the latter on the
above-mentioned parameter, that  is insignificant and  may be
ignored at low energies but is of particular importance at high
energies. Besides, some inferences form the obtained results are
maid.
\end{abstract}

\section{Introduction}
The notion "space-time foam", introduced by J. A. Wheeler about 60
years ago for the description and   investigation of physics at
Planck’s scales (Early Universe) \cite{Wheel1},\cite{mis73}, is
fairly settled. Despite the fact that in the last decade numerous
works have been devoted to physics at Planck’s scales within the
scope of this notion, for example \cite{Gar1}--\cite{dio89}, by
this time still their no clear understanding of the "space-time
foam" as it is.
\\ On the other hand, it is undoubtful that a quantum theory
of  the  Early Universe should be a deformation of the well-known
quantum theory.
\\{\bf The deformation is understood as an extension of a particular
theory by inclusion of one or several additional parameters in
such a way that the initial theory appears in the limiting
transition} \cite{Fadd}.
\\ In his works with the colleagues \cite{shalyt1}--\cite{shalyt9}
the author has put forward one of the possible approaches to
resolution  of a quantum theory at Planck’s scales on the basis of
the density matrix deformation.
 This work demonstrates that  space-time quantum fluctuations,
  in essence generating the space-time foam, may be naturally
described in terms of the deformation parameter $\alpha_{l}$
introduced in  \cite{shalyt1}--\cite{shalyt9}, where $l$ --
measuring scale. Further it is shown that, with the use of quite
natural assumptions, these fluctuations must be allowed for in
Einstein Equations \cite{Einst1} to result in their dependence on
the parameter $\alpha_{l}$, insignificant and negligible at low
energies (i.e. in the limit $l\rightarrow \infty$) but important
at Planck’s scales $l \rightarrow \propto l_P$.
\\ Actually it is revealed that, if the metrics $g_{\mu \nu}$ is measured
at some fixed  energy scale  $E\sim 1/l$ (as is always the case in
real physics), Einstein Equations  are $\alpha_{l}$--deformed, and
the known Einstein Equations \cite{Einst1} appear in the
low-energy limit. However, this aspect  may be ignored in all the
known cases and the corresponding energy ranges because  the scale
$l$ is very distant from $l_P$. Two clear illustrations of the
high-energy $\alpha_{l}$--deformation of Einstein Equations are
given.
\\Some inferences from the results obtained are considered,
in particular for the cosmological term $\Lambda$.
\\This work is a natural continuation of the paper \cite{shalyt-IJMPD}.
In \cite{shalyt-IJMPD} it has been shown that in particular cases
the General Relativity Einstein Equations  may be written in the
$\alpha_{l}$--representation, i.e. they are dependent on the
parameter $\alpha_{l}$. Also, it has been demonstrated that for
the indicated cases one can derive the high-energy (Planck)
$\alpha_{l}$ - deformation of Einstein Equations. Then the
question arises whether Einstein Equations are dependent on
$\alpha_{l}$ in the most general case.
\\Proceeding from the present work, this question may be answered positively.

\section{Quantum Fluctuations of Space-time and High-Energy Deformation}
In accordance with the modern concepts, the space-time foam
\cite{mis73} notion forms the basis for space-time at Planck’s
scales (Big Bang). This object is associated with the quantum
fluctuations generated by uncertainties in measurements of the
fundamental quantities, inducing uncertainties in any distance
measurement. A precise description of the space-time foam is still
lacking along with an adequate quantum gravity theory. But for the
description of quantum fluctuations we have a number of
interesting methods (for example,
\cite{wigner},\cite{found}-\cite{dio89}).
\\ In what follows, we use the terms and symbols from \cite{Ng3}.
Then for the fluctuations $\widetilde{\delta} l$ of the distance
$l$ we have the following estimate:
\begin{equation}\label{Fluct 1}
\widetilde {\delta} l \gsim l_P^{\gamma} l^{1-\gamma},
\end{equation}
where $0\leq\gamma\leq 1$   and  $l_P = (\hbar G/c^3)^{1/2}$ is
the Planck length.
\\ At the present time three principal models associated with different
values of the parameter $\gamma$ are considered:
\\
\\A) $\gamma=1$ that conforms to the initial (canonical) model from \cite{mis73}
\begin{equation}\label{Fluct 1.1}
\widetilde{\delta} l \gsim l_P;
\end{equation}
\\
\\B) $\gamma=2/3$ that conforms to the model \cite{wigner},\cite{Ng3}
compatible with the holographic principle
\cite{Hooft1}--\cite{Bou3}
\begin{equation}\label{Fluct 1.2}
\widetilde{\delta} l \gsim (l l_P^2)^{1/3} = l_P
\left(\frac{l}{l_P}\right)^{1/3};
\end{equation}
\\
\\C) $\gamma=1/2$ - random-walk model \cite{AC}
\cite{dio89}
\begin{equation}\label{Fluct 1.3}
\widetilde{\delta} l \gsim (l l_P)^{1/2} = l_P
\left(\frac{l}{l_P}\right)^{1/2}.
\end{equation}
But, because of the experimental data obtained with the help of
the Hubble Space Telescope \cite{Hubbl1}, a random-walk model C)
may be excluded from consideration (for example, see \cite{Ng8})
and is omitted in this work.
\\
\\ Moreover, in fact it is clear that {\bf at Planck’s scales, i.e. for
\begin{equation}\label{Planck 1.1}
l \rightarrow \propto l_P,
\end{equation}
models A)  are  B)  are coincident}.
\\ Using(\ref{Fluct 1.1})--(\ref{Fluct 1.3}),
we can derive the quantum fluctuations for all the primary
space-time characteristics, specifically for the time
$\widetilde{\delta} t$, energy $\widetilde{\delta} E$, and metrics

$\widetilde{\delta} g_{\mu\nu}$  (formula (10) of \cite{Ng3}):
\begin{equation}\label{Fluct 1.4}
\widetilde{\delta} g_{\mu \nu} \gsim (l_P / l)^{\gamma}.
\end{equation}
It is obvious that all of them are dependent on one and the same
dimensionless parameter $l_P / l$ and on the Planck length $l_P$, i.e. on the fundamental constants.
\\ Note also that in fact this parameter is introduced as
a deformation parameter on going from the well-known quantum
mechanics (QM) to a quantum mechanics with the fundamental length
(QMFL), provided this length $l_{min}$ is on the order of Planck’s
length $l_{min}\propto l_{P}$ , as revealed by the author in the
works written with his colleagues \cite{shalyt1} --\cite{shalyt9}.
Let us recollect in short the central idea of the above-mentioned
works (pp. 1267,1268 in \cite{shalyt2}).
\\ The main object under consideration in this case is the density matrix $\rho$.
We assume that in QMFL   the measuring procedure adopted in QM is
valid being defined by $\rho$. Then
\begin{equation}\label{U6}
Sp[(\rho \widehat{X}^2)-Sp^2(\rho \widehat{X}) ]\geq
l^{2}_{min}>0,
\end{equation}
where $\widehat{X}$ is the coordinate operator. Expression
(\ref{U6}) gives the measuring rule used in QM. However, in the
case considered here, in comparison with QM, the right part of
(\ref{U6}) cannot be done arbitrarily near to zero since it is
limited by $l^{2}_{min}>0$.
 A natural way for studying QMFL is to consider this
theory as a deformation of QM, turning to QM at the low energy
limit (during the expansion of the Universe after the Big Bang).
\\We will consider precisely this option. Here the
following question may be formulated: how should be deformed
density matrix conserving quantum-mechanical measuring rules in
order to obtain self-consistent measuring procedure in QMFL? For
answering to the question we will use the R-procedure. For
starting let us to consider R-procedure both at the Planck's
energy scale and at the low-energy one. At the minimal length
scale $l \approx il_{min}$ where $i$ is a small quantity. Further
$l$ tends to infinity and we obtain for density matrix
\cite{shalyt1}-\cite{shalyt9}:

\begin{equation}\label{Fluct 1.5}
 Sp[\rho l^{2}]-Sp[\rho
l]Sp[\rho l] \simeq l^{2}_{min}\;\; or\;\; Sp[\rho]-Sp^{2}[\rho]
\simeq l^{2}_{min}/l^{2}.
\end{equation}
 Therefore:

 \begin{enumerate}
 \item When $l< \infty$, $Sp[\rho] =
Sp[\rho(l)]$ and
 $Sp[\rho]-Sp^{2}[\rho]>0$. Then, \newline $Sp[\rho]<1$
 that corresponds to the QMFL case.
\item When $l = \infty$, $Sp[\rho]$ does not depend on $l$ and
$Sp[\rho]-Sp^{2}[\rho]\rightarrow 0$. Then, $Sp[\rho]=1$ that
corresponds to the QM case.
\end{enumerate}
The above deformation parameter is as follows:
\begin{equation}\label{Fluct 1.6}
\alpha_{l}= l^{2}_{min}/l^{2}.
\end{equation}
This parameter is variable within the interval
\begin{equation}\label{Fluct 1.7}
0<\alpha_{l}\leq 1/4,
\end{equation}
whereas the density matrix in QMFL  becomes deformed and dependent
on $\alpha_{l}$: $\rho = \rho (\alpha_{l})$, and we get
\begin{equation}\label{U12}
\lim\limits_{\alpha_{l}\rightarrow
0}\rho(\alpha_{l})\rightarrow\rho,
\end{equation}
where $\rho$  -- known density matrix from  QM.
\\ When $l_{min}\propto l_{P}$, it is cleat that
$\alpha_{l} \propto  l^{2}_{P}/l^{2}$ and all the fluctuations
above $\widetilde{\delta} l, \widetilde{\delta} g_{\mu
\nu},\widetilde{\delta} t, \widetilde{\delta} E$ may be expressed
in terms of the deformation parameter  $\alpha_{l}$. For example,
this is the case when the Generalized Uncertainty Principle (GUP)
\cite{Ven1}--\cite{Kempf} is valid
\begin{equation}\label{GUP1.1}
\triangle x\geq\frac{\hbar}{\triangle p}+ \ell^2 \frac{\triangle
p}{\hbar}, \ell^2 =\lambda l_{P}^2 ,
\end{equation}
and $\lambda$ is the model-depended dimensionless numerical factor.
\\Then, as seen in (\ref{GUP1.1}), we have a minimal length
on the order of the Planck length
\begin{equation}\label{GUP2.1}
l_{min}=2\sqrt{\lambda}l_{P}.
\end{equation}
Therefore, we obtain
\begin{equation}\label{GUP3.1}
(\frac{l_{P}}{l})^{2}=\frac{1}{4\lambda}\alpha_{l}
\end{equation}
and the factor  $\frac{1}{4\lambda}$  is introduced into all of
the formula (\ref{Fluct 1.1})--(\ref{Fluct 1.5}) as soon as the
fundamental quantities  involved are expressed in terms of
$\alpha_{l}$. Specifically, the most important formula (\ref{Fluct
1.4}) in this case is of the form \begin{equation}\label{Fluct
1.4A1} \widetilde{\delta} g_{\mu \nu} \gsim (4\lambda)^{-\gamma/2}
\alpha_{l}^{\gamma/2}.
\end{equation}
In what follows we assume that a minimal length in a theory --
$l_{min}$ is existent no matter how it is introduced, from GUP
(\ref{GUP1.1}) or in some other way. Then the parameter
$\alpha_{l}$ (\ref{Fluct 1.6}) is quite naturally brought about
from (\ref{U6}), (\ref{Fluct 1.5}).
\\With the use of this "coordinate system" the above-mentioned models A) and B)
of the  space-time quantum fluctuations may be "unified" as follows:
\\
\\{\bf I. The minimal length $l_{min}$ , similar to cases  A)  and
B), is introduced at Planck’s level
\\
\\$$l_{min}\propto l_{P}$$.
\\
\\II. In both cases fluctuations of the fundamental quantities
may be expressed in terms of the parameter $\alpha_{l}$.
\\
\\III. The principal difference between A) and B) resides in the fact
that in the second case a minimal fluctuation of the length is
dependent on the measuring scale $l$, $(\widetilde{\delta}^{min}
l)=(\widetilde{\delta}^{min} l)[l]$, whereas in the first case it
is completely determined by the minimal length
$\widetilde{\delta}^{min}\approx l_{min}$, being absolute in its
character.
\\
\\IV. As noted above, in the high-energy limit, i.e. for
\begin{equation}\label{Planck 1.2}
l \rightarrow l_{min},
\end{equation}
both models are coincident.}

\section{Quantum Fluctuations and Einstein Equations}
Thus, from the preceding section it follows that in any case we
have minimal fluctuations $\widetilde{\delta}^{min}$ (dependent on
the measuring scale $l$ or on the energy $E\sim 1/l$) for all the
fundamental physical quantities $l,t,E,g_{\mu \nu},...$, expressed
in terms of the parameter  $\alpha_{l}$. Specifically, we have
\begin{equation}\label{Fluct 1.4a}
(\widetilde{\delta}^{min} g_{\mu
\nu})[l]=(\widetilde{\delta}^{min} g_{\mu \nu})[\alpha_{l}]\propto
\alpha_{l}^{\gamma/2}.
\end{equation}
Next we make the only natural assumption
\\{\bf if the metric $g_{\mu \nu}$ in  General
Relativity (GR) is measured at the scale $l$ or, that is the same,
on the scale of the energies $E\sim 1/l$, variation of the metric
$\delta g_{\mu \nu}$ is governed by its fluctuation
$(\widetilde{\delta} g_{\mu \nu})[l]$ and hence it is dependent on
$l$ or, actually, on $\alpha_{l}$}
\\
\\$$\delta g_{\mu \nu}=(\delta g_{\mu \nu})[l]=(\delta g_{\mu
\nu})[\alpha_{l}].$$
\\
In particular, it can’t be arbitrary small as its lower limit is the fixed value
\\
\\$$(\widetilde{\delta}^{min} g_{\mu \nu})[\alpha_{l}]>0.$$
\\
That means
\begin{equation}\label{Fluct 1.4b}
(\delta^{min} g_{\mu \nu})[\alpha_{l}]=\kappa
\alpha_{l}^{\gamma/2},
\end{equation}
where $\kappa>0$ -- some model-dependent factor.
\\ Obviously, we have
\begin{equation}\label{Fluct 1.4c}
\lim\limits_{l\rightarrow \infty}(\delta g_{\mu
\nu})[l]=\lim\limits_{\alpha_{l}\rightarrow 0}(\delta g_{\mu
\nu})[\alpha_{l}]\rightarrow 0.
\end{equation}
From this it follows immediately that in this case variation of
the action of $\delta S_{G}$ in General Relativity \cite{Einst1}
is also dependent on $\alpha_{l}$
\begin{equation}\label{Fluct.act1}
\delta S_{G}=(\delta S_{G})[\alpha_{l}]
\end{equation}
and hence $G_{\mu \nu}\equiv R_{\mu
\nu}-\frac{1}{2}Rg_{\mu \nu}$ is dependent on $\alpha_{l}$ too:
\begin{equation}\label{Fluct.act2}
G^{[\alpha_{l}]}_{\mu \nu}\equiv G_{\mu \nu}[\alpha_{l}].
\end{equation}
Then the knowns {\bf Einstein tensor}
\begin{equation}\label{Fluct.act3A}
\lim\limits_{l\rightarrow \infty}G^{[\alpha_{l}]}_{\mu
\nu}=\lim\limits_{\alpha_{l}\rightarrow 0}G^{[\alpha_{l}]}_{\mu
\nu}\equiv G_{\mu \nu}
\end{equation}
 and {\bf Einstein Equations}  in the vacuum  \begin{equation}\label{Fluct.act3b}
\lim\limits_{l\rightarrow \infty}G^{[\alpha_{l}]}_{\mu
\nu}=\lim\limits_{\alpha_{l}\rightarrow 0}G^{[\alpha_{l}]}_{\mu
\nu}\equiv G_{\mu \nu}=0
\end{equation}
are brought about in the low-energy limit.
\\Naturally, the right side of Einstein Equations \cite{Einst1}
should be dependent on $\alpha_{l}$ as
\begin{equation}\label{Fluct.act3c}
(8\pi T _{\mu \nu}-\Lambda g_{\mu \nu})^{[\alpha_{l}]}\equiv (8\pi
T _{\mu \nu}-\Lambda g_{\mu \nu})[\alpha_{l}].
\end{equation}
Therefore, Einstein Equations with a nonzero right side are of the following form:
\begin{equation}\label{Einst.1}
\lim\limits_{\alpha_{l}\rightarrow 0}G^{[\alpha_{l}]}_{\mu
\nu}=\lim\limits_{\alpha_{l}\rightarrow 0}(8\pi T _{\mu
\nu}-\Lambda g_{\mu \nu})^{[\alpha_{l}]}.
\end{equation}
Of course, at low energies, i.e. for
\begin{equation}\label{Einst.2}
l \gg l_{P}
\end{equation}
\\
\\or,that is the same with a very high accuracy, for
\\
\begin{equation}\label{Einst.3}
\alpha_{l}\approx 0,
\end{equation}
the function of  $\alpha_{l}$  may be disregarded and in this
case, with a very high accuracy, we can obtain the well-known
Einstein Equations
\\
\\$$G^{[\alpha_{l}]}_{\mu
\nu}\approx G_{\mu \nu}=(8\pi T _{\mu \nu}-\Lambda g_{\mu
\nu})\approx(8\pi T _{\mu \nu}-\Lambda g_{\mu
\nu})^{[\alpha_{l}]}.$$
\\
All the scales (energy), at which  Einstein Equations have been
studied until the present time, satisfied
(\ref{Einst.2}),(\ref{Einst.3}), being far away from  the Planck
scale  $l_{P} \propto 10^{-33}cm$, and in fact had no
$\alpha_{l}$--dependence.
\\ But on going to the high-energy limit
\begin{equation}\label{Einst.2.1}
l \rightarrow 2l_{min}\propto l_{P};\alpha_{l}\rightarrow 1/4
\end{equation}
there appears a nontrivial $\alpha_{l}$--deformation of
Einstein  Equations, later referred to as $\alpha$-- deformation
\begin{equation}\label{Einst.Deform}
G^{[\alpha_{l}]}_{\mu \nu}= (8\pi T _{\mu \nu}-\Lambda g_{\mu
\nu})^{[\alpha_{l}]}.
\end{equation}
Note that from \cite{shalyt2} (practically from formula
(\ref{U6}),(\ref{Fluct 1.5})) we have found: with the canonical
measuring procedure (\ref{U6}), the minimal length $l_{min}$ {\bf
is unattainable} and {\bf a minimal measurable length}, denoted as
$l^{measur}_{min}$, is the quantity
\begin{equation}\label{Mesaure1}
l^{measur}_{min}=2l_{min}
\end{equation}
in accordance with (\ref{Einst.2.1}).
\\ Consider two examples of the $\alpha$-- deformation of Einstein Equations.
\\
\\{\bf E1.Phenomenological Markov's Model} \cite{Mark1}.
\\ This example is taken from Section 3 of \cite{shalyt-IJMPD}.
\\ Let us dwell on the work
\cite{Mark1}, where it is assumed that "by the universal decree of
nature a quantity of the material density $\varrho$ is always
bounded by its upper value given by the expression that is
composed of fundamental constants" (\cite{Mark1}, p.214):
\begin{equation}\label{Mark1}
\varrho\leq\varrho_{P}=\frac{c^{5}}{G^{2}\hbar},
\end{equation}
with $\varrho_{P}$ as "Planck's density".
\\ It is clearly seen that, proceeding from the involvement
of the fundamental length on the order of the Planck's
$l_{min}\sim l_{P}$, one can obtain $\varrho_{P}$  (\ref{Mark1})
up to a constant. Indeed, within the scope of GUP (\ref{GUP1.1})
(but not necessarily) we have
 $l_{min}\propto l_{P}$  and then,
 as it has been shown in \cite{shalyt3},
(\ref{GUP1.1}) may be generalized to the corresponding relation of
the pair "energy - time" as follows:
\begin{equation}\label{Mark2}
\Delta t\geq\frac{\hbar}{\triangle E}+\lambda
t_{p}^{2}\frac{\triangle E}{\hbar}.
\end{equation}
This directly suggests the existence of the "minimal time"
$t_{min}\propto t_{P}$ and of the "maximal energy" corresponding
to this minimal time $E_{max}\sim E_{P}$ .
\\Clearly, this maximal energy is associated with some "maximal
mass" $M_{max}$:
\begin{equation}\label{Mark3}
E_{max}=M_{max}c^{2}, M_{max}\sim M_{P}.
\end{equation}
Whence, considering that the existence of a minimal
three-dimensional volume $V_{min}=l^{3}_{min}\sim V_{P}=l^{3}_{P}$
naturally follows from the existence  of $l_{min}\sim l_{P}$, we
immediately arrive at the "maximal density"  $\varrho_{P}$
(\ref{Mark1}) but only within the factor determined by $\lambda$
\begin{equation}\label{Mark4}
\frac{M_{max}}{V_{min}}=\varrho_{max}\sim \varrho_{P}.
\end{equation}
Actually, the quantity
\begin{equation}\label{Mark4.1}
\wp_{\varrho}=\varrho/\varrho_{P}\leq 1
\end{equation}
in \cite{Mark1} is the deformation parameter as it is used to
construct the deformation of Einstein’s equation
(\cite{Mark1},formula (2)):
\begin{equation}\label{Mark5}
R^{\nu}_{\mu}-\frac{1}{2}R\delta^{\nu}_{\mu}=\frac{8\pi
G}{c^{4}}T^{\nu}_{\mu}(1-\wp_{\varrho}^{2})^{n}-\Lambda\wp_{\varrho}^{2n}\delta^{\nu}_{\mu},
\end{equation}
where  $n\geq 1/2$, $T^{\nu}_{\mu}$--energy-momentum tensor,
$\Lambda$-- cosmological  constant. The case of the parameter
$\wp_{\varrho}\ll 1$ or $\varrho\ll \varrho_{P}$ correlates with
the classical Einstein equation, and the case when $\wp_{\varrho}=
1$ --  with the de Sitter Universe. In this way (\ref{Mark5}) may
be considered as $\wp_{\varrho}$-deformation of  the General
Relativity.
\\As it has been noted before, the existence of a maximal
density directly, up to a constant, follows from the existence of
a fundamental length (\ref{Mark1}). It is clear that the
corresponding deformation parameter $\wp_{\varrho}$ (\ref{Mark4.1}
may be obtained from the deformation parameter $\alpha_{x}$
(\ref{Fluct 1.6}). In fact, since $\alpha_{x}=l_{min}^{2}/x^{2}$,
we have
\begin{equation}\label{Mark6}
\alpha_{x}^{3/2}=\frac{l_{min}^{3}}{x^{3}}\sim \frac{V_{min}}{V},
\end{equation}
where $V$ is the three-dimensional volume associated with  the
linear dimension $x$.
\\ As $\alpha_{x}$ may be represented in the form
\cite{shalyt1}--\cite{shalyt9}:
\begin{equation}\label{Mark7}
\alpha_{x}=E^{2}/E^{2}_{max},
\end{equation}
$E_{max}\sim E_{P}$, and $V_{min}\sim V_{P}=l^{3}_{P}$, then from
(\ref{Mark3})--(\ref{Mark4.1}),(\ref{Mark6}),(\ref{Mark7}) we get
\begin{equation}\label{Mark8}
\wp_{\varrho}\sim
\frac{E/V}{E_{max}/V_{min}}=\frac{\varrho}{\varrho_{max}}=\alpha_{x}^{2}.
\end{equation}
Of course, the proportionality factor in (\ref{Mark8}) is model
dependent.
 Specifically, if QMFL is related to GUP, this factor is depending on
 $\lambda$ (\ref{GUP1.1}).
 But the deformation parameters $\wp_{\varrho}$ and
 $\alpha$ are differing considerably: the limiting value $\wp_{\varrho}=1$
 is obviously associated with singularity, whereas originally
 (by the approach involving the density matrix deformation
 \cite{shalyt2}--\cite{shalyt4},\cite{shalyt9})
no consideration has been given to the deformation parameter
 $\alpha=1$ associated with singularity,(formula (\ref{Mesaure1}))).
\\ So, $\wp_{\varrho}$-deformation of  the  General
Relativity \cite{Mark1} may be interpreted as
$\alpha$-deformation.
\\
\\ {\bf E2.Spherically-symmetric horizon spaces} \cite{Padm13}.
\\ As shown in \cite{Padm13}, the Einstein Equation for horizon
in this case may be written as a thermodynamic identity (the first
principle of  thermodynamics): (\cite{Padm13}, formula (119))
\begin{equation}\label{GT12}
   \underbrace{\frac{{{\hbar}} cf'(a)}{4\pi}}_{\displaystyle{k_BT}}
    \ \underbrace{\frac{c^3}{G{{\hbar}}}d\left( \frac{1}{ 4} 4\pi a^2 \right)}_{
    \displaystyle{dS}}
  \ \underbrace{-\ \frac{1}{2}\frac{c^4 da}{G}}_{
    \displaystyle{-dE}}
 = \underbrace{P d \left( \frac{4\pi}{ 3}  a^3 \right)  }_{
    \displaystyle{P\, dV}},
\end{equation}
 where a static, spherically symmetric horizon in
space-time is described by the metric
\begin{equation}\label{GT9}
ds^2 = -f(r) c^2 dt^2 + f^{-1}(r) dr^2 + r^2 d\Omega^2,
\end{equation}
and the horizon location will be given by simple zero of the
function $f(r)$ ($f(a)=0$, $f'(a)\ne 0$) at $r=a$.( Here $r=a$ is
the radius of a sphere.) And $P = T^{r}_{r}$ is the trace of the
momentum-energy tensor and radial pressure.
\\ In Sections 5 and 6 of \cite{shalyt-IJMPD} first the Einstein Equations on horizon (\ref{GT12}) have been written
 in terms of the parameter $\alpha_{a}$, next the high-energy ($\alpha_{a}\rightarrow 1/4$), $\alpha_{a}$
-- deformation of these equations has been derived in two different cases:
equilibrium and nonequilibrium thermodynamics.
\\ The latter case is distinguished from the first one by the dynamic cosmological term dependent
on $\alpha_{a}$, appearing with the corresponding factor in the
right side of high-energy deformed  (\ref{GT12}) as follows:
\begin{equation}\label{Dynamics1}
\Lambda\equiv \Lambda[\alpha_{a}].
\end{equation}

\section{Comments and Conclusion}
In this way we can conclude that
\\
\\C1) with inclusion of the space-time quantum fluctuations
(e.g., in the form of (\ref{Fluct 1.1}) or (\ref{Fluct 1.2}),
we can naturally assume that in the most general case of Einstein
Equations there is a dependence on the small dimensionless
parameter $\alpha_{l}$, infinitesimal at normal energies to be
neglected but important at high energies which are on the order of
the Planck energy.
\\
\\C2) The parameter $\alpha_{l}$  is a deformation parameter
on going from the well-known quantum theory to a quantum theory of
the Early Universe (Planck’s scales)and hence the above-mentioned
dependence of  Einstein Equations on this parameter may be
considered as $\alpha_{l}$ -- deformation of the General
Relativity whose well-known, i.e. canonical, Einstein Equations
are brought about in the low-energy limiting transition.
\\
\\ The foregoing results are rather important for better
understanding and investigation of the cosmological term
$\Lambda$, especially in view of the Dark Energy Problem
\cite{Dar1.1}--\cite{Dar1.5}.
\\ In principle, they may be used to answer
the question whether $\Lambda=const$ or $\Lambda=\Lambda(t)$ is a
time-variable quantity.
\\Despite the fact that the works taking $\Lambda$
as $\Lambda(t)$, i.e. as a dynamic quantity, are numerous(for
example, \cite{varyingL.1}-- \cite{varyingL.4}) quite forceful
arguments are given against this point of view (for example,
\cite{Dolgov}).
\\Indeed, according to the General Relativity, the cosmological term $\Lambda$
has been considered constant $\Lambda=const$ as, due to the
Bianchi identities \cite{Einst1},
\begin{equation}\label{bianchi}
 \nabla^{\mu} G_\mu{}_\nu = 0.
\end{equation}
But in this work it has been demonstrated that, actually, Bianchi
identities (\ref{bianchi}) are introduced at the low-energy limit
only
\begin{equation}\label{bianchi2}
 \lim\limits_{\alpha_{l}\rightarrow 0}\nabla^{\mu} G^{[\alpha_{l}]}_{\mu \nu}=\nabla^{\mu} G_\mu{}_\nu = 0.
\end{equation}
Because of this, the really measured cosmological term $\Lambda$
in fact is dynamic $\Lambda=\Lambda[\alpha_{l}(t)]$, practically
invariable in the modern epoch, i.e. at low energies, due to  slow
variations of the deformation parameter $\alpha_{l}(t)$  at low
energies and due to its very small value.
\\ In the works \cite{shalyt-entropy2}--\cite{Shalyt-Entropy2012}
a behavior of the term $\Lambda$ has been studied reasoning from
$\alpha_{l}(t)$  on the assumption that it is dynamic, similar to
the case proven in \cite{shalyt-entropy2} GUP   for the pair of
conjugate variables $(\Lambda,V)$, where $V$ is the space-time
volume, as with the holographic principle applied to  the whole
Universe  \cite{Sussk1}. The main difference of these two cases is
in the leading order of expansion $\Lambda[\alpha]$ in terms of
$\alpha$. In the first case it is the second
\begin{equation}\label{DE9.2}
\Lambda^{GUP}(\alpha) \propto (\alpha^{2}
+\eta_{1}\alpha^{3}+...)\Lambda_{p},
\end{equation}
whereas in the second case it is the first
\begin{equation}\label{DE9}
\Lambda^{Hol}(\alpha) \propto (\alpha
+\xi_{1}\alpha^{2}+...)\Lambda_{p},
\end{equation}
where $\Lambda_{p}=\Lambda_{\alpha\rightarrow 1/4}$ --
cosmological term at Planck’s scales.
\\ As $\Lambda^{Hol}$  is practically coincident
with the experimental value of the cosmological term
$\Lambda_{exper}$, a holographic model is preferable -- model B)
of Section 2 developed for quantum fluctuations is supported
experimentally.
\\In conclusion, let us state some important problems of the particular concern:
\\
\\A) What is the way to derive, in the most general case and in the explicit form,
the  high-energy ($\alpha_{l}\rightarrow 1/4$) $\alpha_{l}$ -
representation or, that is the same, the  high-energy $\alpha_{l}$
- deformation of Einstein Equations?
\\
\\B) Provided the foregoing representation is derived, is it possible to have
its logical series expansion in terms of  $\alpha_{l}$? Note that
we must allow for the following:  $\alpha_{l}$ may be considered
continuous with a high accuracy only at low energies. Obviously,
at high energies it is discrete as the length $l$ is comparable to
the minimal length $l\propto l_{min}$, i.e. in fact to the
Planck’s length $l\propto l_{P}$.
\\As noted in point IV of Section 2, on approximation of the Planck energies,
models (A) and (B) for the space-time fluctuations are practically
coincident. Because of this, we can raise the following questions:
\\
\\$C_{1}$) Is there some {\bf "critical measure"} or {\bf "critical index"}
$\gamma_{crit}$:$\gamma=2/3 <\gamma_{crit} < \gamma=1$ -- minimal
bound, beginning from which models (A) and (B) are practically
identical at high energies, between the coefficients $\gamma=2/3$
and $\gamma=1$ in formulae (\ref{Fluct 1.2})  and (\ref{Fluct
1.2})? If such a {\bf "critical index"} exists, what is it like?
This may be of great importance for answering the question that
concerns the {\bf "phase transition"}, i.e. the minimal energies,
beginning from which one should  take into account the
quantum-gravitational effects.
\\
\\ Another but similar problem:
\\
\\$C_{2}$) concerns a minimal bound for $\alpha_{l}$
(denoted by  $\alpha^{crit}_{l}=l^{2}_{min}/l_{crit}^{2}$), above
which models (A)  and  (B) actually result in the same physics. It
is clear that the problem at hand is associated with derivation of
the corresponding energy: $E_{crit}\sim 1/ l_{crit}$.
\\
\\ And, finally,
\\(D) it is interesting how the  high-energy $\alpha_{l}$ - deformation
of Einstein Equations is related to the adequate selection of a
model for the space-time foam. Is it representing a set of micro
worm holes(for example, \cite{Gar1}--\cite{Gar4}), micro black
holes \cite{Scard1}-- \cite{Scard3} or something else?
\\The author is planning to answer these questions, at least some of them, in his future works.


\begin{thebibliography}{}

\bibitem{Wheel1}
J. A. Wheeler,{\it Geometrodynamics} (Academic Press, New York and
London, 1962).

\bibitem{mis73}
C. W. Misner, K. S. Thorne, and J. A. Wheeler, {\it Gravitation}
(Freeman, San Francisco, 1973).

\bibitem{Gar1}
Remo Garattini, \emph{Int. J. Mod. Phys. D} {\bf 4} (2002)  635.

\bibitem{Gar2}
Remo Garattini,\emph{Entropy} {\bf 2} (2000) 26.

\bibitem{Gar3}
Remo Garattini, \emph{Nucl.Phys.Proc.Suppl.} {\bf 88} (2000) 297.

\bibitem{Gar4}
Remo Garattini, \emph{Phys.Lett.B} {\bf 459}  (1999) 461.

\bibitem{Scard1}
Fabio Scardigli,\emph{Class.Quant.Grav.} {\bf 14} (1997) 1781.

\bibitem{Scard2}
Fabio Scardigli,\emph{Phys.Lett.B} {\bf 452}  (1999) 39.

\bibitem{Scard3}
Fabio Scardigli, \emph{Nucl.Phys.Proc.Suppl.} {\bf 88} (2000) 291.

\bibitem{Garay1}
Luis J. Garay,\emph{Phys.Rev. D} {\bf 58} (1998) 124015.

\bibitem{Garay2}
 Luis J. Garay, \emph{Phys.Rev.Lett.}  {\bf 80} (1998) 2508.

\bibitem{found}
Y. J. Ng and H. van Dam, \emph{Found. Phys.} {\bf 30} (2000) 795.

\bibitem{stfoam}
Y. J. Ng, \emph{Int. J. Mod. Phys. D} {\bf 11} (2002) 1585.

\bibitem{Ng3}
Y. J. Ng, \emph{Mod.Phys.Lett.A}  {\bf 18} (2003) 1073

\bibitem{Ng4}
Y. J. Ng, gr-qc/0401015.

\bibitem{Ng5}
Y. J. Ng, H. van Dam,\emph{Int.J.Mod.Phys.A} {\bf 20} (2005) 1328.

\bibitem{Ng6}
W.A. Christiansen, Y. Jack Ng, H. van Dam, \emph{Phys.Rev.Lett.}
{\bf 96} (2006) 051301

\bibitem{Ng7}
Y. Jack Ng, \emph{Phys.Lett.B} {\bf 657} (2007)  10.

\bibitem{Ng8}
Y. Jack Ng, \emph{AIP Conf.Proc.} {\bf 1115} (2009)  74.

\bibitem{Ng9}
A. Wayne  Christiansen, David J. E. Floyd, Y. Jack Ng, Eric S.
Perlman,  \emph{Phys.Rev.D}  {\bf 83} (2011) 084003.

\bibitem{AC}
G. Amelino-Camelia, \emph{Nature} {\bf 398}  (1999) 216.

\bibitem{dio89}
L. Diosi and B. Lukacs, \emph{Phys. Lett.} {\bf A142} (1989) 331.

\bibitem{Fadd}
 L.Faddeev,  \emph{Priroda} {\bf 5} (1989) 11.


\bibitem{shalyt1}
A.E. Shalyt-Margolin  and J.G. Suarez,  \emph{gr-qc/0302119}.
%
%
\bibitem{shalyt2}
A.E. Shalyt-Margolin  and J.G. Suarez,  \emph{Intern. Journ. Mod.
Phys D}  {\bf 12} (2003) 1265.
%
%
\bibitem{shalyt3}
A.E. Shalyt-Margolin  and A.Ya. Tregubovich,
 \emph{Mod. Phys.Lett. A}  {\bf 19} (2004) 71.
%
%
\bibitem{shalyt4}
A.E. Shalyt-Margolin, \emph{Mod. Phys. Lett. A}  {\bf 19} (2004)
391.
%
%
\bibitem{shalyt5}
A.E. Shalyt-Margolin, \emph{Mod. Phys. Lett. A} {\bf 19} (2004)
2037.
%
%
\bibitem{shalyt6}
A.E. Shalyt-Margolin, \emph{Intern. Journ. Mod.Phys D}  {\bf 13}
(2004)  853.
%
%
\bibitem{shalyt7}
A.E. Shalyt-Margolin, \emph{Intern.Journ.Mod.Phys.A}  {\bf 20}
(2005) 4951.
%
%
\bibitem{shalyt8}
A.E. Shalyt-Margolin and  V.I. Strazhev, The Density Matrix
Deformation in Quantum and Statistical Mechanics in Early
Universe. In \emph{Proc. Sixth International Symposium "Frontiers
of Fundamental and Computational Physics"},ed. B.G. Sidharth
(Springer, 2006) p.~131.
%
%
\bibitem{shalyt9}
A.E. Shalyt-Margolin, The Density matrix deformation in physics of
the early universe and some of its implications. In \emph{Quantum
Cosmology Research Trends},ed. A. Reimer (Horizons in World
Physics. {\bf 246}, Nova Science Publishers, Inc., Hauppauge,
NY,2005) p.~49.
%
%
\bibitem{Einst1}
Robert. M. Wald,  {\it General Relativity}, (The University
Chicago Press Chicago and London 1984).
%
%
\bibitem{wigner}
E.P. Wigner, \emph{Rev. Mod. Phys.} {\bf 29}  (1957) 255.

\bibitem{Hooft1}
G. 'T. Hooft, \emph{gr-qc/9310026}.
%
%
\bibitem{Hooft2}
G. 'T. Hooft, \emph{hep-th/0003004};
%
%
\bibitem{Sussk}
L.Susskind, \emph{J. Math. Phys}  {\bf 36} (1995) 6377.
%
%
\bibitem{Bou1}
R. Bousso, \emph{Rev. Mod. Phys} {\bf 74} (2002)  825.
%
%
\bibitem{Bou3}
R. Bousso, \emph{JHEP}  {\bf 07} (1999)
 004.
%
%
\bibitem{Hubbl1}
E. S. Perlman et al., Astron. J.{\bf  124}, 2401–2412 (2002)
%
%

\bibitem{Ven1}
G. A. Veneziano, {\it Europhys.Lett} {\bf 2} (1986) 199.
%
%

\bibitem{Ven2}
D. Amati, M. Ciafaloni,  and G. A. Veneziano,  {\it Phys.Lett.B}
{\bf 216} (1989) 41.

\bibitem{Ven3}
E.Witten, {\it Phys.Today} {\bf 49} (1996) 24.
%
%
\bibitem{GUPg1}
R.~J.~Adler,D.~I. ~Santiago, \emph{Mod. Phys. Lett. A} {\bf 14},
1371 (1999).
%
%
\bibitem{Ahl1}
D.V.Ahluwalia,  \emph{Phys.Lett} {\bf  A275}  (2000)  31.

\bibitem{Ahl}
 D.V.Ahluwalia, \emph{Mod.Phys.Lett}  {\bf  A17} (2002) 1135.
%
%
\bibitem{Magg1}
M. Maggiore, \emph{Phys.Lett B}  {\bf 319}  (1993) 83.
%
%
\bibitem{Kempf}
A. Kempf, G. Mangano and R.B. Mann, \emph{Phys.Rev.D}  {\bf 52}
(1995) 1108.
%
%
\bibitem{Mark1}
M.A. Markov, \emph{Pis'ma v ZHETF}  {\bf 36 } (1982) 214.
%
%
\bibitem{shalyt-IJMPD}
A.E. Shalyt-Margolin, \emph{Intern. J.  Mod. Phys. D} {\bf 21}
(2012) 1250013.
%
%
\bibitem{Padm13}
T. Padmanabhan, \emph{Rep. Prog. Phys.} {\bf 73} (2010) 046901,
arXiv:0911.5004.
%
%
\bibitem{Dar1.1}
S.Perlmutter et al., {\it Astrophys. J} {\bf 517} (1999) 565.
%
%
\bibitem{Dar1.2}
A. G. Riess et al.,  {\it Astron. J} {\bf 116} (1998) 1009.
%
%
\bibitem{Dar1.3}
 A. G.Riess et al.,  {\it Astron. J}  {\bf 117} (1999) 707.
%
%
\bibitem{Dar1.4}
V. Sahni and A. A. Starobinsky, {\it Int. J. Mod. Phys. D} {\bf 9}
(2000) 373.
%
%
\bibitem{Dar1.5}
 S. M. Carroll,  {\it Living Rev. Rel}  {\bf 4} (2001)1.
%
%
\bibitem{varyingL.1}
O. Bertolami, {\it N. Cim. B} {\bf 93}  (1986) 36.
%
%
\bibitem{varyingL.2}
 J.C. Carvalho,J.A.S Lima and I. Waga, {\it Phys. Rev. D} {\bf 46} (1992)
 2404.
%
%
\bibitem{varyingL.3}
L.P. Chimento and D. Pavon,  {\it Gen. Rel. Grav.} {\bf 30} (1998)
643.
%
%
\bibitem{varyingL.4}
T. Harco and M.K. Mak, {\it Gen. Rel. Grav.}  {\bf 31} (1999) 849.
%
%
\bibitem{Dolgov}
 A.D. Dolgov, \emph{Phys.Atom.Nucl} {\bf 71} (2008) 651.
%
%
\bibitem{shalyt-entropy2}
A.E. Shalyt-Margolin, \emph{Entropy} {\bf 12}  (2010)  932.
%
%
\bibitem{Shalyt-ijtmp1}
A.E. Shalyt-Margolin,  \emph{Intern. J.  Theor. Math. Phys.} {\bf
1} (2011)  1.
%
%
\bibitem{Shalyt-Entropy2012}
A.E. Shalyt-Margolin, \emph{Entropy} {\bf 14}  (2012)  2143.
%
%
\bibitem{Sussk1}
W. Fischler and L. Susskind, \emph{hep-th/9806039}.
%
%
\end{thebibliography}
\end{document}